\documentclass[]{iopart}
\usepackage{graphicx}
\usepackage{lastpage}
\usepackage{amssymb}

\newcommand{\ket}[1]{|#1 \rangle}

\begin{document}
\title{Single photon quantum non-demolition in the presence of inhomogeneous broadening.}

\author{Andrew D.~Greentree$^1$, R. G. Beausoleil$^2$, L. C. L. Hollenberg$^1$
W. J. Munro$^{3,4}$, Kae Nemoto$^4$, S. Prawer$^1$, and T. P.
Spiller$^3$}
\address{$^1$Centre for Quantum Computer Technology, School of
Physics, The University of Melbourne, Melbourne, Victoria 3010,
Australia.}
\address{$^2$Hewlett-Packard Laboratories, 1501 Page Mill Rd., Palo Alto, CA  94304-1123, USA}
\address{$^3$Hewlett-Packard Laboratories, Filton Road, Stoke Gifford, Bristol
BS34 8QZ, United Kingdom.}
\address{$^4$National Institute of Informatics, 2-1-2 Hitotsubashi,
Chiyoda-ku, Tokyo 101-8430, Japan.}

\ead{andrew.greentree@ph.unimelb.edu.au}
\date{\today}

\begin{abstract}
Electromagnetically induced transparency (EIT) has been often
proposed for generating nonlinear optical effects at the single
photon level; in particular, as a means to effect a quantum
non-demolition measurement of a single photon field. Previous
treatments have usually considered homogeneously broadened samples, but
realisations in any medium will have to contend with inhomogeneous
broadening. Here we reappraise an earlier scheme [Munro \textit{et
al.} Phys. Rev. A \textbf{71}, 033819 (2005)] with respect to
inhomogeneities and show an alternative mode of operation that is
preferred in an inhomogeneous environment. We further show the
implications of these results on a potential implementation in diamond
containing nitrogen-vacancy colour centres. Our modelling
shows that single mode waveguide structures of length
$200~\mu\mathrm{m}$ in single-crystal diamond containing a dilute ensemble of NV$^-$ of only 200 centres are
sufficient for quantum non-demolition measurements using EIT-based
weak nonlinear interactions.  
\end{abstract}



\section{Introduction}
The importance of quantum mechanics to modern technology is
indisputable. However, what remains to complete the `quantum
revolution' \cite{bib:DowlingQP02} is the exploitation of
\emph{coherent} quantum mechanics in technological devices as well
as the incoherent quantum mechanics responsible for, e.g.,
transistor electronics. Systems of strongly interacting photons and
atoms have long been convenient systems for probing coherent quantum
mechanics through the field of quantum optics.

Of all the effects between coherently prepared atoms and light,
electromagnetically induced transparency (EIT) is often promoted as
an important building block for physics and device applications,
because EIT allows the possibility of large optical nonlinearities
accompanied by complete transparency \cite{bib:HarrisPRL1990}. EIT
is a coherent quantum phenomenon whereby the absorptive and
dispersive properties of a three (or more) state system can be
tailored by using applied electromagnetic fields, and we discuss its
properties more fully below. First observed by Boller, Imamo\u{g}lu
and Harris \cite{bib:BollerPRL1991}, some of the proposed
applications for EIT include magnetometry \cite{bib:ScullyPRL1992},
high-efficiency UV generation
\cite{bib:MerriamOL1999,bib:DormanPRA2000}, photonic switches
\cite{bib:HarrisPRL1998}, and optical \cite{bib:Beausoleil2004} quantum gates, and light storage in first-in first
out (FIFO) networks \cite{bib:TuckerJLT2005,bib:MatskoPRA2007}.
Although the medium of choice is usually a vapour cell (e.g. Rb
\cite{bib:ZibrovPRL1995}), future technology may be more easily realised with solid-state media.  EIT has also been studied in solids
\cite{bib:HamOC1997,bib:WeiJOB1999,bib:HemmerOptLett2001,bib:KuznetsovaPRA2002,bib:SantoriOptExp2006,bib:SantoriPRL2006,bib:AlegrePRB2007},
magneto-optical traps \cite{bib:Chen1998} and Bose-Einstein
Condensates \cite{bib:HauNature1999}.

Here, we concentrate on the possibility of using the lossless Kerr
nonlinearity assocatiated with EIT for realising a quantum
non-demolition (QND) measurement.  QND via the cross-Kerr effect between two distinct optical modes was
originally proposed by Imoto, Haus and Yamamoto
\cite{bib:ImotoPRA1985}, and invoking the EIT induced Giant Kerr
nonlinearity is a popular suggestion for realising a $\pi/2$ phase
shift for such a measurement.  The idea that such a QND gate could
be realised by weak nonlinearities, such as are routinely found in
EIT systems, \emph{without} a full $\pi/2$ phase shift induced on
the detection beam, was introduced in Ref.~\cite{bib:MunroPRA2005}.
In this system, the weak nonlinearity is effectively enhanced by the
presence of a strong probe beam.  However the earlier proposal did
not consider all of the limitations of realistic systems, and in
particular did not consider the effect of inhomogeneous broadening
of the EIT medium: we do so here. Although we are concentrating on
QND measurements, it has been shown that QND measurements effected
by weak nonlinearities can act as a primitive for other quantum
gates, \cite{bib:NemotoPLA2005} and this directly leads to the Qubus
\cite{bib:NemotoPRL2004,bib:MunroNJP2005,bib:MunroJOptB2005,bib:SpillerNJP2006}
and related schemes for quantum repeaters \cite{bib:LoockPRL2006}
and cluster-state generation \cite{bib:LouisNJP2007}. The results
presented here should be equally applicable to these, and other
EIT-related schemes.

\begin{figure}[tb!]
\includegraphics[width=0.9\columnwidth,clip]{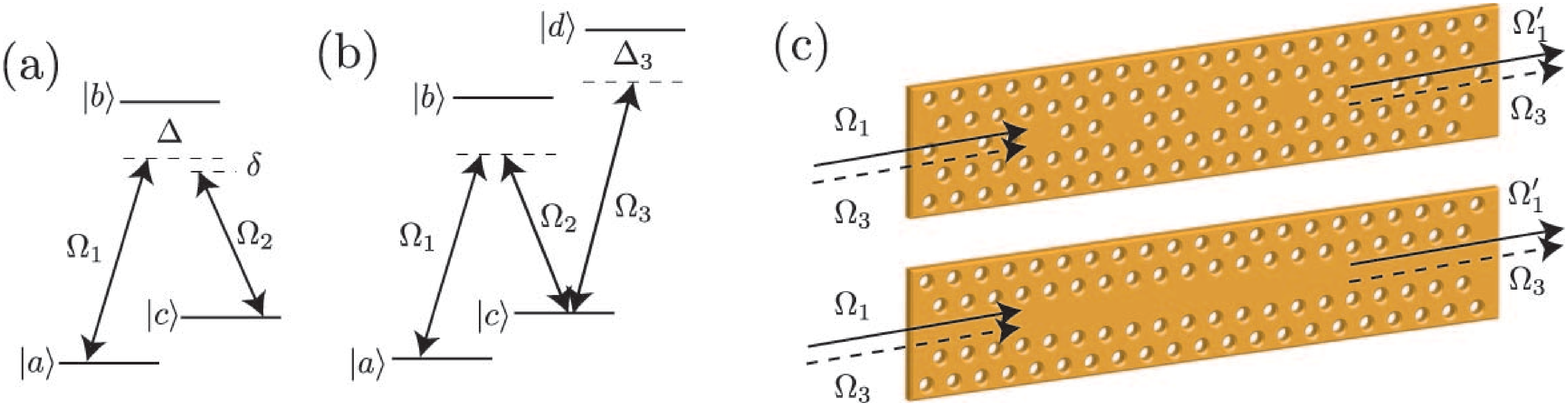}
\caption{\label{fig:3LA} (a) Schematic of the three-state $\Lambda$
system under consideration.  States $|a\rangle$ and $|c\rangle$ are
ground (meta-stable) states, and have no direct spontaneous emission
pathways.  $|b\rangle$ is an excited state, which decays to
$|a\rangle$ and $|c\rangle$ with equal probabilities. The
$|a\rangle-|b\rangle$ transition is driven by field 1, with detuning
$\Delta$ and Rabi frequency $\Omega_1$ (pump).  The
$|b\rangle-|c\rangle$ transition is driven by field 2 (probe) with
detuning $\Delta+\delta$ (i.e. $\delta$ is the shift from the mutual
detuning $\Delta$) and Rabi frequency $\Omega_2$. (b) Schematic of
four-state system in the $N$ configuration, where the $\Lambda$
system is modified by an interaction with field 3 driving
(off-resonantly) the $|c\rangle-|d\rangle$ transition with Rabi
frequency $\Omega_3$ and detuning $\Delta_3$. To first order, we
expect the $|c\rangle-|d\rangle$ transition to perturb the EIT in
the $|a\rangle-|b\rangle-|c\rangle$ system via the usual light
shift, which plays a role equivalent to the two-photon detuning in
the three-state scheme. (c) Proposed configurations for realising
the QND measurement.  Fields 1 and 3 enter a single-mode diamond
waveguide constructed from a photonic bandgap material with tailored group velocity for field 3 and of order $100~\mu\mathrm{m}$ long. Two illustrations of potential structures are shown, the lower follows Krauss \cite{bib:KraussJPD2007}, and the upper is via coupled cavity arrays in the style of Altug and Vu\v{c}kovi\'{c} \cite{bib:AltugAPL2005}
Field 2 (not shown) illuminates from the side.  The unknown field, field 3,
propagates unperturbed, but field 1 is phase shifted by the number
of photons in field 3, and this phase shift can be read out using
heterodyne methods (not shown).}
\end{figure}

EIT is a well-known mechanism for generating optical nonlinearities
without loss \cite{bib:HarrisPRL1990}, for a recent review see
Ref.~\cite{bib:FleischhauerRMP2005}.  The typical (and minimal)
system in which to observe EIT is a three-state system in the
$\Lambda$ configuration with two driving fields, depicted in
Fig.~\ref{fig:3LA}(a) with states labelled $|a\rangle$, $|b\rangle$
and $|c\rangle$. EIT is a manifestation of quantum interference in
an atomic system: Considering field 1 as a `pump' and field 2 as a
`probe', absorption of the probe is suppressed due to the coherence
induced by the pump. The coherence gives rise to interference
between the dressed states on the $|a\rangle-|b\rangle$ transition,
and hence a dip in the probe absorption.  Because the absorption dip
is due to a coherent two-photon resonance (the resonant two-photon
transition from $|a\rangle$ to $|c\rangle$) it can, in principle, be
much less than the optical linewidth of the $|b\rangle-|c\rangle$
transition, limited only by the decoherence on the
$|a\rangle-|c\rangle$ transition.

\begin{figure}[tb!]
\includegraphics[width=10.5cm,clip]{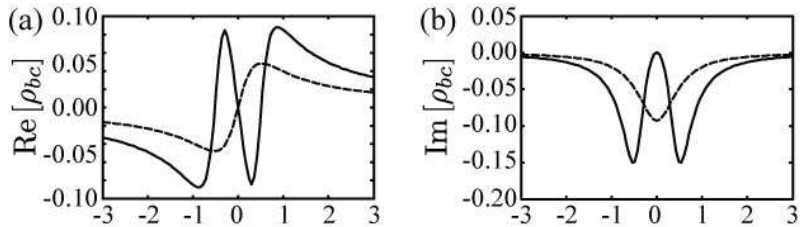}
\caption{\label{fig:ReviewEIT} Typical dispersion (a) and absorption
(b) curves associated with EIT in a $\Lambda$ system (solid lines)
compared to those of a two-state system with equal probe couplings
(dashed lines). The narrow EIT resonance is associated with linear
dispersion at $\delta=0$. The steep slope makes it easy to perturb
the EIT resonance and gives rise to the Giant Kerr effect.}
\end{figure}

Although transparency is the eponymous feature of EIT, of more
practical interest is the steep and controllable dispersion curve
that is associated with the transparency point.  A typical example
of the dispersive and absorptive spectra associated with EIT are
shown in Fig.~\ref{fig:ReviewEIT}. At the transparency point
(two-photon resonance between the ground states) there is a linear
dispersion. This feature has led to dramatic demonstrations of
ultra-slow group velocity light
\cite{bib:HauNature1999,bib:KashPRL1999,bib:TurukhinPRL2002} and is
the basis for the Giant Kerr nonlinearity \cite{bib:SchmidtOL1996}
in the $N$ configuration of Fig.~\ref{fig:3LA} (b). Considering the
$N$ system as a $\Lambda$ system perturbed by an off-resonant
transition, one can see that the detuned off-resonant third field
induces a small light shift to $|c\rangle$. Although this is a small
effect, because of the steepness of the dispersion, a large effect
on the EIT resonance is observed.  This property was proposed for
achieving photonic blockade in cavity QED systems
\cite{bib:ImamogluPRL1997}, and has undergone extensive theoretical investigation (e.g.
Refs.~\cite{bib:RebicJOB1999,bib:WernerPRA2000,bib:GreentreeJOB2000,bib:HartmannQP2006,bib:BermelPRA2006})
and recently observed \cite{bib:BirnbaumNat2005} (although in the
two-state, rather than four-state configuration). When field 3 is
resonant, an absorptive, rather than dispersive, nonlinearity is
observed, which has been studied theoretically
\cite{bib:HarrisPRL1998} and experimentally
\cite{bib:YanPRA2001,bib:EchanizPRA2001}, and will not be treated
here.  Neither do we consider the many related atomic
configurations, e.g. the Tripod \cite{bib:PaspalakisJMO2002},
extended N \cite{bib:Zubairy2002}, Chain-$\Lambda$
\cite{bib:GreentreePRA2003}, and  $M$-scheme
\cite{bib:RebicPRA2004}, although they all offer potential
improvements over the conventional $N$ scheme.

One important detail for realising nonlinear interactions in the $N$
system is that of group velocity matching.  In a travelling wave
geometry to realise the $N$ system of Fig.~\ref{fig:3LA} (b), we
require the pulses that describe fields 1 and 3 to be temporally
coincident for the maximum cross-Kerr interaction (the classical
pump field 2 can be assumed to be derived from a large uniform
field, and so is exempt from this criterion).  Because field 1 is
travelling under EIT conditions, it will be propagating with
extremely slow group velocity: field 3 is not.  We see that the
amount of mutual interaction would therefore be expected to be
limited by the temporal walk-off of the two pulses.  There have been
many suggestions in the literature to counter this effect (e.g.
Refs.~\cite{bib:LukinPRL2000,bib:WangPRL2006}) which invoke varying
levels of complexity of the interaction medium. It is also possible
to control group velocity by modification of the medium, e.g. by
using tailored photonic bandgap structures
\cite{bib:VlasovPRE1999,bib:VlasovNature2005}. Because we are mainly
interested in solid-state implementations, we will assume that the
system  is embedded in a photonic-crystal structure where the group
velocity as seen by field 3 is tuned by the structure to balance the
EIT induced group velocity seen by field 1, and we therefore will
not treat this detail further.

Although much of our treatment in this paper will be system
independent, it is important to note that a major motivation for
performing this reappraisal of weak nonlinear gates is the
availability of a new material for observing optical EIT: diamond
containing the negatively charged Nitrogen-Vacancy colour centre
(NV).  This material has shown quite remarkable results, including
single photon generation (e.g. \cite{bib:BeveratosEPJD2002}),
room-temperature Rabi oscillations \cite{bib:JelezkoPRL2004}, and
spin-spin coupling
\cite{bib:JelezkoPRL2004b,bib:GaebelNatPhys2006,bib:HansonPRL2006,bib:DuttScience2007}.
EIT has been demonstrated in NV diamond in the rf
\cite{bib:WeiJOB1999} and optical regimes
\cite{bib:SantoriOptExp2006,bib:SantoriPRL2006}. The intrinsic
properties of NV centres can also be modified by fabricating optical
structures directly in ultra-nanocrystalline \cite{bib:WangAPL2007}
or single crystal diamond
\cite{bib:OliveroAdvMat2005,bib:GreentreeJPCM2006,bib:Fairchild}, by
growth on preexisting optical structures \cite{bib:RabeauAPL2005},
and also by Stark shifting which has been demonstrated on bulk
samples \cite{bib:Redman} and spectrally resolved centres
\cite{bib:TamaratPRL2006,bib:TamaratCM}.  The time is therefore ripe
to examine NV diamond for the goal of optical quantum information
processing.

In the next section, we discuss EIT and the properties of coherently
driven $\Lambda$ systems.  By including inhomogeneous broadening and
expanding about the EIT point we are able to derive analytical
results for the absorption and dispersion without the usual
assumption of weak probe and strong pump fields.  We are also able
to determine optimal ratios for pump and probe Rabi frequencies.  In
Section \ref{sect:4LA} we show the results for cross-Kerr
nonlinearities in the four-state N system, and in
Section~\ref{sect:Gates} we present our main results, which use the
results from the preceeding sections to design structures based on
diamond containing the Nitrogen-Vacancy colour centre that should be
sufficient to realise a number-discriminating Quantum Non-Demolition
operation.

\section{Three-state $\Lambda$ system}
A schematic of our model three-state system is shown in
Fig.~\ref{fig:3LA}.  Following Shore \cite{bib:ShoreBook}, we write
down the Hamiltonian under the rotating wave approximation in matrix
form with state ordering $|a\rangle,|b\rangle,|c\rangle$
\begin{eqnarray}
\mathcal{H} = \hbar \left(\Delta \sigma_{bb} + \delta \sigma_{cc}
+\Omega_1 \sigma_{ba} + \Omega_2 \sigma_{cb} + h.c.\right),
\label{eq:Ham}
\end{eqnarray}
where field 1 (field 2) drives the $|a\rangle-|b\rangle$
($|b\rangle-|c\rangle$) transition with Rabi frequency $\Omega_1$
($\Omega_2$) and detuning $\Delta_1$ ($\Delta_2$), and $\sigma_{ij}
= |i\rangle\langle j|$. As we are operating near the two-photon
resonance, we set $\Delta_1 = \Delta$ and $\Delta_2 = \Delta+\delta$
where $\Delta$ is the mutual detuning, and $\delta$ is the detuning
from two-photon resonance. Spontaneous emission at rate $\Gamma$ is
from $|b\rangle$ to $|a\rangle$ and $|c\rangle$ with equal
probabilities. We treat inhomogeneous broadening by
considering a distribution of $\Delta$, i.e. the mutual detunings.  As we are
only considering inhomogeneity on the excited state distribution, the \emph{two-photon} detuning, $\delta$, does not vary.  Note that although we will use
terminology such as `pump'  and `probe', we will usually make no
assumption about the relative strength of these fields.  In this way
our analysis is analogous to the cases treated by Wielandy and Gaeta
on EIT in the strong pump regime \cite{bib:WielandyPRA1998} or the
parametric EIT regime \cite{bib:MullerPRA2000}.

One way to proceed in gaining insight on the EIT problem set out in
Eq.~\ref{eq:Ham} is to construct the master equation and determine
the steady state solution. This can be expressed as
\begin{eqnarray}
\dot{\rho} = -\frac{i}{\hbar} [\mathcal{H},\rho] + \sum_j \Gamma_j
\mathcal{L}[B_j,\rho], \label{eq:ME}
\end{eqnarray}
where we have introduced the usual density matrix, $\rho$ and the
Liouvillian super-operator, $\mathcal{L}[B,\rho]$ which describes
the effect of the generalised decoherence channel $B$ with rate
$\Gamma \geq 0$ on the density matrix, and is summed over all
decoherence channels. The Liouvillian operators are defined
\begin{eqnarray}
\mathcal{L}[B,\rho] \equiv  B\rho B^{\dag} - \frac{1}{2}
\left(B^{\dag}B\rho + \rho B^{\dag}B\right).
\end{eqnarray}
In our case,we restrict ourselves to the case that the system is limited by
spontaneous emission, and that the decoherence between the ground
states can be neglected.  This approximation is warranted because of
the very long ground state decoherence rates in important systems of
interest (e.g. diamond containing the negatively charged
Nitrogen-Vacancy colour centre \cite{bib:GaebelNatPhys2006}, or
rubidium vapour cells \cite{bib:KashPRL1999}) but does limit the
\emph{minimum} Rabi frequencies that can be applied to be larger
than this decoherence rate.  So here, the $B$ will be the one-way
(spontaneous emission) transitions from $|b\rangle$ to either
$|a\rangle$ ($\sigma_{ab}$) or $|c\rangle$ ($\sigma_{cb}$) with
rate, $\Gamma/2$.

To analyse Eq.~\ref{eq:ME} in the steady state, we convert the
master equation into superoperator form, and write
\begin{eqnarray}
\dot{\vec{\rho}} = \left(-i \mathcal{P} +
\mathcal{L}\right)\vec{\rho},
\end{eqnarray}
where $\vec{\rho}$ is the vector obtained by writing out the density
matrix elements, and $\mathcal{P}$ and $\mathcal{L}$ are the
superoperators describing Hamiltonian and decoherence processes
respectively.

\begin{figure}[tb]
\includegraphics[width=105mm,clip]{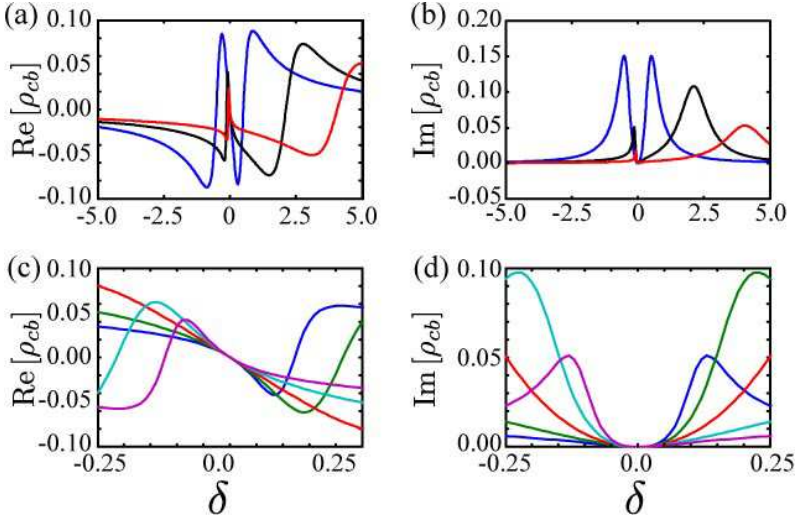}
\caption{\label{fig:InHomEIT1} EIT in for homogeneously broadened
systems. (a) $\mathrm{Re}(\rho_{cb})$ as function of $\delta$ for
the mutual detunings of $\Delta = 0,~2\Gamma,~4\Gamma$ (blue, black
and red lines respectively) to illustrate the effect of moving away
from resonance. (b) Absorption [$\mathrm{Im}(\rho_{cb}$] under the
same conditions. Again note that all of the EIT transparency windows
overlap. (c) Closeup in the vicinity of $\delta = 0$, showing
$\mathrm{Re}(\rho_{cb})$ this time for $\Delta =
-2\Gamma,~-\Gamma,~0,~\Gamma,~2\Gamma$ (blue, green, red, cyan and
magenta lines respectively) (d) Analogous spectra for
$\mathrm{Im}(\rho_{cb})$.}
\end{figure}

\begin{figure}[tb]
\includegraphics[width=105mm,clip]{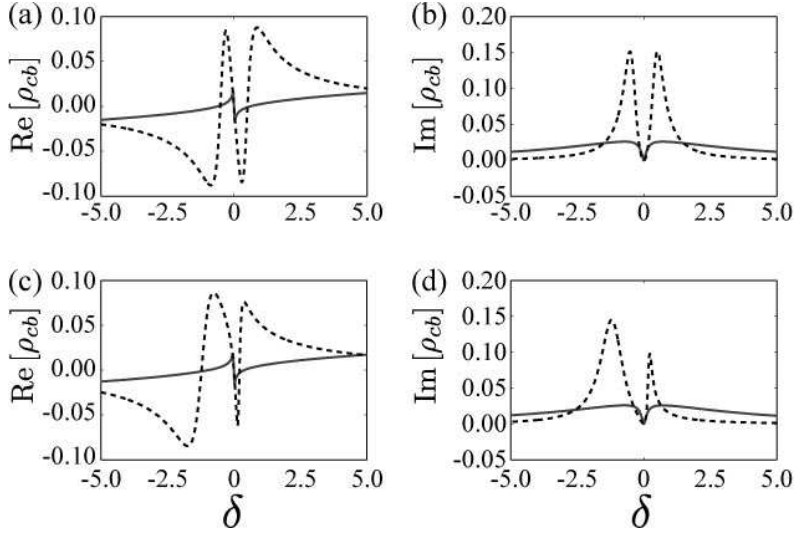}
\caption{\label{fig:InHomEIT2} Comparison of EIT traces as a
function of the detuning from the two-photon resonance condition,
$\delta$, for one-photon resonance, (a) real part, (b) imaginary
part, and finite mutual detuning, $\Delta = 2\Omega$, (c) real part
and (d) imaginary part. In each case the dashed line is the
single-atom result, whilst the solid lines are the average over the
inhomogeneous line, with $\gamma = 10\Gamma$. Although the overall
lineshapes vary with $\Delta$ and inhomogeneity, the $\delta = 0$
responses are equivalent to second order.}
\end{figure}

In addition to the master equations, we also include the effect of
inhomogeneous broadening. Inhomogeneous broadening has been treated
previously in the context of Doppler-broadened EIT in vapour cells
(e.g. \cite{bib:GeaBanalochePRA1995,bib:VemuriPRA1996}), and is
treated as a Gaussian distribution of the absolute energy of
$|b\rangle$, which in turn is manifested as a variation in $\Delta$
across the sample, i.e. there is a probability distribution of
detunings
\begin{eqnarray}
P(\Delta) = \frac{1}{\sqrt{2\pi \gamma^2}}
\exp\left[-\frac{\left(\Delta-\Delta_0\right)^2}{2\gamma^2}\right],
\end{eqnarray}
where $\Delta_0 = \int_{-\infty}^{\infty} \Delta P(\Delta) d\Delta$
is the mean mutual detuning with respect to the inhomogeneous
linewidth, which has standard deviation $\gamma$.
Fig.~\ref{fig:InHomEIT1} shows spectra with the real and imaginary
parts of the $\ket{b}-\ket{c}$ coherence, $\rho_{bc}$ (proportional
to the probe dispersion and absorption respectively) with $\Omega_1
= \Gamma/2$ and $\Omega_2 = \Gamma/10$ to illustrate the effect of
inhomogeneous broadening on the resonant EIT profile.
Figs.~\ref{fig:InHomEIT1} (a) and (b) show the absorption and
dispersion curves for $\Delta$ from 0 to positive values showing the
effect of increasing $\Delta$. Note that spectra of the same colour
go together.  Although the overall spectra are quite different, in
the vicinity $\delta=0$ they are all locally similar.  This is
clearer in Figs.~\ref{fig:InHomEIT1} (c) and (d) which shows spectra
from large negative to positive $\Delta$ superimposed, showing the
local similarity strongly. Figs.~\ref{fig:InHomEIT1} (e) and (f)
show the effect over the whole inhomogeneous linewidth with $\gamma
= 10\Gamma$ (solid blue line) compared with the a sample with equal
total population but only homogeneously broadened sample (black
dashed line).  Note that these plots may equally well be interpreted
as classic hole-burning spectra \cite{bib:AllenBook}. The
self-similarity of the EIT traces is not dependent upon the mutual
detuning, as is illustrated in Fig.~\ref{fig:InHomEIT2} which
compares $\Delta_0 = 0$ with $\Delta_0 = -\Gamma$.

The self-similarity of the EIT profiles about $\delta=0$ may be
understood by considering $\rho_{bc}$ in the steady state. Setting
$\dot{\vec{\rho}}=0$, we determine the null-space which gives the
non-trivial steady state solution for $\vec{\rho}^{(ss)}$. In
general the analytical results are somewhat complicated. By
confining our interest to the region around $\delta=0$, i.e. in the
vicinity of the two-photon resonance, we may perform a series
solution for $\vec{\rho}^{(ss)}$ in powers of $\delta$, which yields
to first order
\begin{eqnarray}
\rho_{aa} = \frac{\Omega_2^2}{\Omega_1^2+\Omega_2^2} - \frac{2 \Omega_1^2 \Omega_2^2 \Delta}{\left(\Omega_1^2 + \Omega_2^2\right)^3}\delta, \\
\rho_{ab} = \frac{\Omega_1\Omega_2^2}{\left(\Omega_1^2 + \Omega_2^2\right)^2}\delta = \rho_{ba}^{\ast},\\
\rho_{ac} =-\frac{\Omega_1\Omega_2}{\Omega_1^2+\Omega_2^2} + \frac{\Omega_1\Omega_2\left[\Delta\left(\Omega_1^2 - \Omega_2^2\right) - i \Gamma\left(\Omega_1^2 + \Omega_2^2\right)\right]}{\left(\Omega_1^2 + \Omega_2^2\right)^3}\delta = \rho_{ca}^{\ast}, \\
\rho_{bb} =   0, \\
\rho_{cb} =-\frac{\Omega_1^2 \Omega_2}{\left(\Omega_1^2 + \Omega_2^2\right)^2}\delta = \rho_{bc}^{\ast}, \label{eq:RhocbFirstOrder} \\
\rho_{cc} = \frac{\Omega_1^2}{\Omega_1^2 + \Omega_2^2} + \frac{2
\Omega_1^2\Omega_2^2 \Delta}{\left(\Omega_1^2 +
\Omega_2^2\right)^3}\delta.
\end{eqnarray}
Cursory inspection of these steady state results provides some very
important properties of the EIT condition. Modelling of inhomogenous
broadening was to be by varying $\Delta$ over the ensemble, however
assuming field 1 is the pump and field 2 the probe, then the
parameter of interest for EIT is $\rho_{cb}$. From the above, we see
the well-known linear dependance with detuning expected for an EIT
resonance, which has \emph{no} dependence on $\Delta$, indicating
that inhomogeneous broadening will affect neither the absorption nor
dispersion seen by the probe field to first order in the probe
detuning (but to all orders in the mutual detuning). This is an
important result, highlighting that EIT is extremely robust to
inhomogeneous broadening in the excited state.  To see effects due
to the inhomogenous broadening of the line, we will need to go to
higher orders in $\delta$.

The total coherence is obtained by integrating over the
inhomogeneous linewidth, so we have
\begin{eqnarray}
\varrho_{cb} = \int_{-\infty}^{\infty} P(\Delta) \rho_{cb} ~d\Delta,
\end{eqnarray}
where we have introduced $\varrho_{cb}$ as the coherence integrated
over the inhomogeneous line.  As $\int_{-\infty}^{\infty} P(\Delta)
d\Delta = 1$, to first order in $\delta$, we have trivially that
$\varrho_{cb} = \rho_{cb}$.  To see effects due to the inhomogeneous
line, we must go to second order in $\delta$, where we have
\begin{eqnarray}
\rho_{cb}^{(2)} = \frac{\Omega_1^2\Omega_2}{(\Omega_1^2 +
\Omega_2^2)^2}\left[-\delta
 + \frac{\left(\Omega_1^2 -3\Omega_2^2\right) \Delta \delta^2}{(\Omega_1^2 +
 \Omega_2^2)^2}
 + i \frac{\Gamma \delta^2}{(\Omega_1^2 + \Omega_2^2)}\right] + \mathcal{O}[\delta]^3.
\nonumber \\
\label{eq:rhocbSecondOrder}
\end{eqnarray}
Performing the integration over the inhomogeneous line yields
\begin{eqnarray}
\varrho_{cb}^{(2)} &=& \int_{-\infty}^{\infty} P(\Delta)
\rho_{cb}^{(2)} d\Delta, \\
 &=& \frac{\Omega_1^2\Omega_2}{(\Omega_1^2 + \Omega_2^2)^2}\left[-\delta
 + \frac{\left(\Omega_1^2 - 3\Omega_2^2\right) \Delta_0 \delta^2}{(\Omega_1^2 +
\Omega_2^2)^2} + i\frac{\Gamma \delta^2}{(\Omega_1^2 + \Omega_2^2)}\right].
\end{eqnarray}

\begin{figure}[tb!]
\includegraphics[width=0.8\columnwidth,clip]{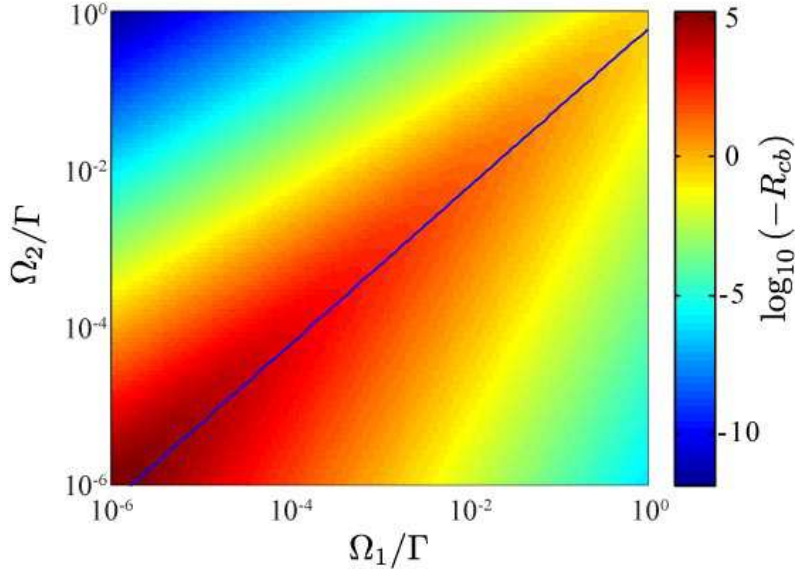}
\caption{\label{fig:gradientvsO1O2} First order probe dispersion,
plotted as $\log_{10}(-R_{cb})$ as a function of $\Omega_1$ (pump)
and $\Omega_2$ (probe) in the vicinity of two-photon resonance in a
three-state system. Note that the dispersion as seen by the probe
will be largest when $\Omega_1 = \sqrt{3} \Omega_2$, and goes as
$R_{cb} = - 3\sqrt{3}/(16\Omega_1)$ in this limit.  This optimal
solution for $R$ is indicated by the solid line.}
\end{figure}

Regions with high dispersion are associated with regions of high
nonlinearity, and hence will inform us in our search for optimal
working points for our gates.  The probe dispersion is defined
\begin{eqnarray}
R_{cb} \equiv \frac{\partial \mathrm{Re} [\rho_{cb}]}{\partial
\delta},
\end{eqnarray}
and using the first order solution in eq.~\ref{eq:RhocbFirstOrder}
we obtain
\begin{eqnarray}
R_{cb} = -\frac{\Omega_1^2 \Omega_2}{\left(\Omega_1^2 + \Omega_2^2
\right)^2}. \label{eq:dispersion}
\end{eqnarray}
Note that if we set $\Omega_1 = \sqrt{3}\Omega_2$, then the second
order correction to the refractive index in
Eq.~\ref{eq:rhocbSecondOrder} is nulled, and we must therefore go to
third order in $\delta$ (or higher) to observe terms depending
explicitly on the inhomogeneous broadening. This result also carries
through to the dispersion, i.e. that the gradient is largest when
$\Omega_1$ and $\Omega_2$ are smallest and maximised when $\Omega_1
= \sqrt{3}\Omega_2$.  A graph showing $\log_{10}(-R)$ as a function
of $\Omega_1$ and $\Omega_2$ is shown in
Fig.~\ref{fig:gradientvsO1O2}, along with a line showing the maximal
dispersion at $\Omega_1 = \sqrt{3}\Omega_2$.  These results may be
useful to optimise slow and stopped light experiments, however for
the purposes of QND measurements (as will be seen in
Section~\ref{sect:Gates}) the susceptibility is simply maximised for
minimum possible $\Omega_1$.

\begin{figure}[tb!]
\includegraphics[width=0.8\columnwidth,clip]{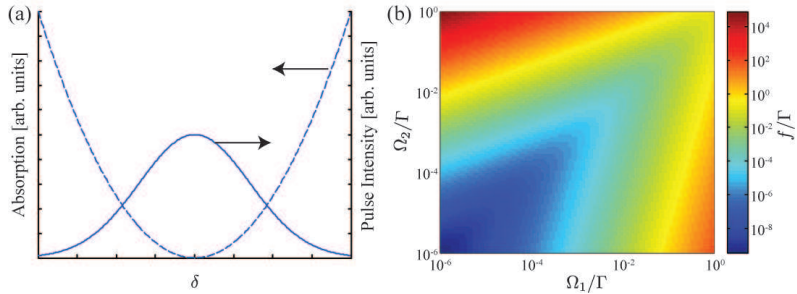}
\caption{\label{fig:AbsorptionSchematic} (a) Schematic showing the
absorption seen by an optical pulse with finite bandwidth.  To
second order, the EIT window has absorption that is quadratic in
$\delta$, where a transform limited pulse is a Gaussian in $\delta$,
so the residual absorption per unit length seen by a pulse
travelling through an EIT window will be found by taking the overlap
integral of the pulse and absorption. (b) Log base 10 of the maximum
bandwidth in units of $\Gamma$ that will attain $\mathcal{A} = 1\%$
as a function of $\Omega_1/\Gamma$ and $\Omega_2/\Gamma$.  Note that
although the bandwidths can be quite small, they are still of order
the Rabi frequencies of the fields, which accords with our intuition
about the width of the EIT window.  The smallest allowable bandwidth
coincides with the largest dispersions, again as one would expect.}
\end{figure}

To explore the absorption we take the imaginary part of the solution
for $\rho_{cb}$, which is to third order in $\delta$
\begin{eqnarray}
A = \frac{\Omega_1^2 \Omega_2 \Gamma}{(\Omega_1^2 +
\Omega_2^2)^3}\delta^2 -
2\Delta\Gamma\frac{\Omega_1^2\Omega_2(\Omega_1^2 -
\Omega_2^2)}{(\Omega_1^2 + \Omega_2^2)^5}\delta^3 +
\mathcal{O}[\delta^4].
\end{eqnarray}
Ignoring the third order correction, here we see the familiar
quadratic dependance with respect to detuning, and as presaged
above, there is no contribution to the absorption from $\Delta$.  We
can infer the bandwidth of the EIT medium, by considering the effect
of the EIT window on a pulse with finite bandwidth.
Fig.~\ref{fig:AbsorptionSchematic}(a) shows schematically a
transform limited Gaussian pulse propagating through the quadratic
EIT window.  If we assume that the pulse is defined by some spectral
width $f$, centered around the EIT window at $\delta=0$, with the
functional form $\exp[-\delta^2/(2f^2)]/\sqrt{2\pi f^2}$, then to
determine the total (single atom) absorption we integrate the pulse
over the EIT window, i.e.
\begin{eqnarray}
\mathcal{A} &=& \int_{-\infty}^{\infty}
\frac{\exp\left[-\delta^2/(2f^2)\right]}{\sqrt{2\pi f^2}}
\frac{\Omega_1^2 \Omega_2 \Gamma}{(\Omega_1^2 +
\Omega_2^2)^3} \delta^2 d\delta, \\
 &=& \frac{\Omega_1^2 \Omega_2 \Gamma f^2}{(\Omega_1^2 +
 \Omega_2^2)^3}, \label{eq:AbsBandWidth}
\end{eqnarray}
and this quantity will prove essential in determining bandwidth
requirements in the design of practical nonlinear gates.  To explore
this, we calculate in Fig.~\ref{fig:AbsorptionSchematic}(b) the
bandwidth in units of the spontaneous emission, $f/\Gamma$ as a
function of $\Omega_1/\Gamma$ and $\Omega_2/\Gamma$ that will
achieve a $\mathcal{A} = 1\%$. Such analyses as these allow us to
determine gate speeds for QND measurements and will be exploited in
Section~\ref{sect:Gates}.

Finally, we comment on the group velocity seen under conditions of
EIT.  Group velocity reduction is one of the most dramatic
consequences of EIT and is observable in the usual pump-probe
arrangement (e.g. \cite{bib:HauNature1999,bib:KashPRL1999}). Perhaps
surprisingly, large changes in the group velocity between fields
(group velocity mismatch) can actually lead to a strong
\emph{reduction} in effective coupling.  As mentioned previously, we
propose group velocity engineering as the solution to this mismatch,
but it is essential to understand the group velocity under
conditions of EIT in order to specify the propagation properties of
the unknown field.

The group velocity of a probe field can be determined
using Eq.~\ref{eq:dispersion}, so we have
\begin{eqnarray}
v_{g_{ab}} &=& \frac{c}{\eta + \frac{\omega_1}{2} \frac{\partial
\chi_{ab}}{\partial \Delta}}, \\
\end{eqnarray}
where $\eta$ is the bulk refractive index, and we have introduced the susceptibility
\begin{eqnarray}
\chi_{ab} = \frac{2\pi \mathcal{N}}{\varepsilon_0 \varepsilon_R \hbar}\frac{\mu_{ab}^2}{\Omega_1}\rho_{ab}, 
\end{eqnarray}
and so
\begin{eqnarray}
\frac{\partial \chi_{ab}}{\partial \delta} &=& \frac{2\pi \mathcal{N}}{\varepsilon_0 \varepsilon_R \hbar}\frac{\mu_{ab}^2}{\Omega_1}R_{ab}, \\
	&=& \frac{2\pi \mathcal{N}}{\varepsilon_0 \varepsilon_R \hbar}\mu_{ab}^2 \frac{\Omega_2^2}{\left(\Omega_1^2 + \Omega_2^2\right)^2}, 
\end{eqnarray}
where $\mathcal{N}$ is the number density of atoms.  So in the limit that the group velocity reduction is large, we have
\begin{eqnarray}
v_{g_{ab}} = \frac{c \varepsilon_0 \varepsilon_R \hbar}{\pi\omega \mathcal{N} \mu_{ab}^2}\frac{\left(\Omega_1^2 + \Omega_2^2\right)^2}{\Omega_2^2}.
\end{eqnarray}

Ultimately we will be interested in the group velocity associated with quantised fields, rather than the semiclassical form above.  Making the substitution $\tilde{\Omega}_1 \sqrt{n_1} = \Omega_1$ where $n_1$ is the number of photons in the mode, gives the group velocity seen by a mode with $n_1$ photons of
\begin{eqnarray}
v_{g_{ab}} = \frac{c \varepsilon_0 \varepsilon_R \hbar}{\pi \omega \mathcal{N} \mu_{ab}^2}\frac{\left(\tilde{\Omega}_1^2 n_1 + \Omega_2^2\right)^2}{\Omega_2^2}.  \label{eq:QuantumGroupVelocity}
\end{eqnarray}

These results will be used in Section~\ref{sect:Gates}, especially
with quantisation of the probe field.  In general one would seek the largest possible
$v_g$ that enables effective coupling, so as to minimise the
reduction in the group velocity on the unknown signal which is not
travelling under conditions of EIT reduced group velocity.  Also the
group velocity dispersion which will manifest with uncertainty in the number of photons in the probe field is also a potential source of
error and should be minimsed.  This implies a rule of thumb, that we should seek operation in the limit
of large $n_1$ to minimise the relative variation in $v_g$.

\section{Four-state $N$ system}\label{sect:4LA}

The four-state $N$ scheme is one level structure that clearly shows
a cross-Kerr effect, and is illustrated in Fig.~\ref{fig:3LA}(b).
There is much freedom to choose which fields correspond to pump,
probe and driving, and all appear to have been treated in the
literature in various places. For concreteness, we will treat the
$N$ system as a $\Lambda$ perturbed by an off-resonant transition.
Furthermore, we will be considering the effect of the
$|b\rangle-|d\rangle$ transition (hence field 3) on field 1, so in
this section, the parameters of interest will be $\rho_{ab}$ and
$\varrho_{ab}$. For this case we set the operating point for the
$\Lambda$ system as $\delta=0$. Under these conditions, the
Hamiltonian is
\begin{eqnarray}
\mathcal{H} = \hbar \left(\Delta \sigma_{bb} + \Delta_3 \sigma_{dd}
+\Omega_1 \sigma_{ba} + \Omega_2 \sigma_{cb} + \Omega_3 \sigma_ {dc}
+ h.c.\right), \label{eq:Ham4x4}
\end{eqnarray}
where the (in general unknown) Rabi frequency of field 3 is
$\Omega_3$ and it is detuned from the $|b\rangle-|d\rangle$
transition by $\Delta_3 \gg \Omega_3$.

In the case that the $|c\rangle-|d\rangle$ transition is only
homogenously broadened, we may treat the effect of field 3 on the
$|a\rangle-|b\rangle-|c\rangle$ system quite simply.  Our treatment
here follows and extends Ref.~\cite{bib:Beausoleil2004}. The effect
of a field 3 on the $|c\rangle-|d\rangle$ transition can be seen as
an off-resonant light-shift, which in turn perturbs the
$|a\rangle-|b\rangle-|c\rangle$ $\Lambda$ EIT.  The strength of the
light shift can be directly equated with the $-\delta$ (the shift
will be in the opposite direction) from the previous analysis, i.e.
\begin{eqnarray}
\delta = -\frac{\Omega_3^2}{\Delta_3},
\end{eqnarray}
in the limit that $\Delta_3 \gg \Omega_3$. Recalling that the
residual population in $|d\rangle$ is a source of error, we will
assume that this population must be kept below some threshold,
$\epsilon$, i.e.
\begin{eqnarray}
\rho_{dd} = \left(\frac{\Omega_3}{\Delta_3}\right)^2 \leq \epsilon.
\label{eq:rhoddineq}
\end{eqnarray}
Using the previous result for the steady state of $\rho_{ab}$, and
substituting for $\delta$, we have
\begin{eqnarray}
\rho_{ab} = -\frac{\Omega_1 \Omega_2^2}{\left(\Omega_1^2 +
\Omega_2^2\right)^2}\frac{\Omega_3^2}{\Delta_3}.
\label{eq:LightShiftCoherence}
\end{eqnarray}

The Hamiltonian of Eq.~\ref{eq:Ham4x4} can also be attacked using
the superoperator approach.  We assume that all decay from
$|d\rangle$ is to $|c\rangle$ at rate $\Gamma$, and keeping the
other terms the same from the $\Lambda$ analysis, except $\Delta=0$.
We can obtain the steady state response quite easily, but for
clarity, we only report the coherence on the $|a\rangle-|b\rangle$
transition, which is
\begin{eqnarray}
\rho_{ab} = -\frac{2 \Omega_1 \Omega_2^2 \Omega_3^2}{(2 \Delta_3 + i
\Gamma)(\Omega_1^2 + \Omega_2^2)^2} + \mathcal{O}[\Omega_3]^4,
\label{eq:4StateRhoab}
\end{eqnarray}
which is qualitatively very similar to the previous result, but with
the inclusion of an extra absorption term.

\begin{figure}[tb!]
\includegraphics[width=0.8\columnwidth,clip]{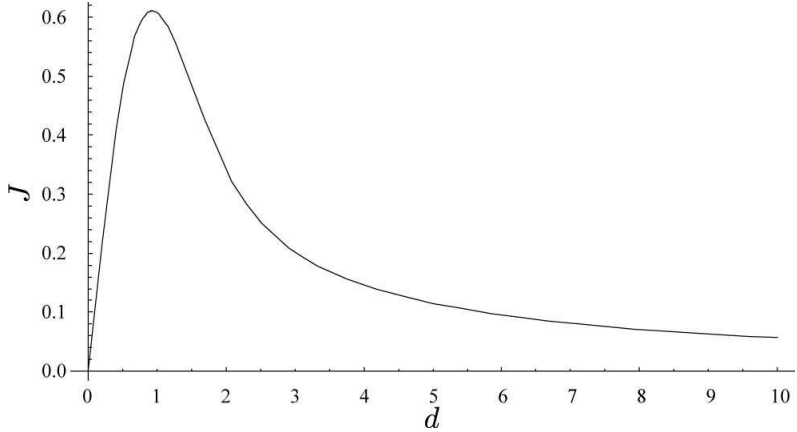}
\caption{\label{fig:OptimalMeanDetuning} Graph showing optimal mean
detuning from resonance in units of the inhomogeneous linewidth.
Here we have taken our figure of merit to be $J(d)\equiv
\exp(-d^2){\rm erfi}(d)$ which is the term in $\varrho_{ab}$ which
depends explicitly on $d = \bar{\Delta}/(2\gamma)$.  Note that the
results for $d \lesssim 1$ will not be relevant for our discussion
because the condition $\Omega_3 \ll \Delta_3$ may be unsatisfied,
and there may also be significant \emph{resonant} one-photon
absorptions of field 3}
\end{figure}

To move to the case of a finite inhomogeneous linewidth, we first
recall that we may safely ignore linewidth to first order on the
$\Lambda$ system, however, we cannot do so on the
$|b\rangle-|d\rangle$ transition, as $\Delta_3$ appears directly in
the coherence. Therefore, using the light-shifted treatment from
Eq.~\ref{eq:LightShiftCoherence} we write down the ensemble
coherence as
\begin{eqnarray}
\varrho_{ab} &=& \int_{-\infty}^{\infty} P(\Delta_3) \rho_{ab} d\Delta_3, \\
 &=& -\frac{\Omega_1 \Omega_2^2 \Omega_3^2}{\sqrt{2 \pi \gamma^2} \left(\Omega_1^2 +
\Omega_2^2\right)^2} \int_{-\infty}^{\infty}
\frac{\exp[-(\Delta_3-\Delta_0)^2/(2 \gamma^2)]}{\Delta_3}d\Delta_3,
\\
 &=& -\frac{\sqrt{\pi} \Omega_1 \Omega_2^2 \Omega_3^2}{\sqrt{2 \gamma^2} \left(\Omega_1^2 +
\Omega_2^2\right)^2} \exp\left(-\frac{\Delta_0^2}{2 \gamma^2}\right)
{\rm erfi}\left(\frac{\Delta_0}{\sqrt{2 \gamma^2}}\right),
\end{eqnarray}
where ${\rm erfi}(x) = {\rm erf}(ix)/i$ is the imaginary error
function. Note that although we have treated the inhomogeneous
linewidths for $|d\rangle$ as equivalent to that of $|b\rangle$, the
analysis is practically unchanged if we took a more general case. It
is instructive to examine the behaviour of just the exponential and
imaginary error function terms, to determine the optimal ratio of
mean detuning to inhomogeneous linewidth.  Setting $d =
\Delta_0/\sqrt{2\gamma^2}$ as the mean detuning in units of the
linewidth, and $J(d) \equiv \exp(-d^2){\rm erfi}(d)$ in
Fig.~\ref{fig:OptimalMeanDetuning} we plot $J(d)$ vs $d$, as for
constant $\gamma$, all the other terms are constant. The maximum of
$J$ can be seen as being just before $d=1$, however, care must be
taken in this limit, and there will still be appreciable absorption
of field 3 in this limit (as there will be on-resonant atoms).
However, detuning by 5 inhomogenous linewidths (i.e. $d=5$) only has
the effect of reducing the effective $\varrho_{ab}$ by about one
sixth compared to a homogeneously broadened sample at such a
detuning: the larger effect is the penalty in having to go to such
large detunings to avoid the line (i.e. for a homogenously broadened
sample, one could work closer to resonance).

An alternative transition to be considered for the readout is the
$|c\rangle-|d\rangle$ transition.  The coherence associated with
this can also be determined and is
\begin{eqnarray}
\rho_{cd} = -\frac{2 \Omega_1^2 \Omega_3}{(2\Delta_3 +
i\Gamma)(\Omega_1^2 + \Omega_2^2)} + \mathcal{O}[\Omega_3]^3.
\label{eq:4StateRhocd}
\end{eqnarray}
Integrating this over the inhomogeneity on the $|c\rangle-|d\rangle$
transition does not yield analytic solutions, however by replacing
the denominator by the large detuning approximation (i.e. the term
$(2\Delta_3 + i\Gamma)/2$ is replaced by $\Delta_3$, we get
\begin{eqnarray}
\varrho_{cd} = -\frac{\sqrt{\pi}\Omega_1^2\Omega_3}
 {\sqrt{2\gamma^2}(\Omega_1^2 + \Omega_2^2)}
 \exp \left(-\frac{\Delta_0}{2\gamma^2}\right)
 \textrm{erfi} \left(\frac{\Delta_0}{\sqrt{2 \gamma^2}}\right).
\end{eqnarray}

For completeness, we also report the coherence on the
$|c\rangle-|b\rangle$ transition, which is
\begin{eqnarray}
\rho_{cb} = \frac{2 \Omega_1^2 \Omega_2 \Omega_3^2}{(2\Delta_3 -
i\Gamma)(\Omega_1^2 + \Omega_2^2)^2}.
\end{eqnarray}
As the response here is clearly analogous to that of $\rho_{ab}$ we
will not repeat any further discussion of using this transition for
monitoring any probe state, except to note that changing between
$|a\rangle-|b\rangle$ and $|c\rangle - |b\rangle$ may perhaps be
useful for reasons of experimental convenience in certain
implementations, but otherwise would appear to hold no benefits.

It is interesting to note that neither Eq.~\ref{eq:4StateRhoab} nor
\ref{eq:4StateRhocd} exhibit self-Kerr effects, with the first self
nonlinearities for $\rho_{ab}$ appearing at fourth order, and for
$\rho_{cd}$ appearing at third order.  The presence of Kerr terms
would hamper state discrimination in nonlinear gates (see next
section) \cite{bib:Rohde2008,bib:Kok2007}.  One should note that the
canonical method for generating self-Kerr terms in EIT media is to
allow a field to interact with more than one transition (e.g.
\cite{bib:ImamogluPRL1997,bib:Zubairy2002,bib:GreentreePRA2003})
which effectively converts the cross-Kerr nonlinearity into a
self-Kerr nonlinearity.  The suppression of self-Kerr terms is
desirable, as they give rise to pulse distortion which can limit the
effectiveness of any gate based on nonlinear interactions.

Finally, we comment further on the populations in the excited
states.  In Eq.~\ref{eq:rhoddineq} we presented a simple two state
argument for the population in state $|d\rangle$, which was viewed
as a potential source of error.  We now examine the full solutions
for $\rho_{bb}$ and $\rho_{dd}$, which turn out to be qualitatively
similar to the analysis based on perturbing the EIT structure by the
extra transition.

Starting with $\rho_{dd}$, we find that the expansion to third order
in $\Omega_3$ yields
\begin{eqnarray}
\rho_{dd} = \frac{\Omega_1^2}{\Omega_1^2 + \Omega_2^2}
\frac{\Omega_3^2}{\Delta_3^2 + (\Gamma/2)^2} +
\mathcal{O}[\Omega_3]^4
\end{eqnarray}
This result can be immediately interpreted as the usual,
off-resonant population from Eq.~\ref{eq:rhoddineq} (with
spontaneous emission explicitly included), scaled by a factor due to
the diminished population in $|c\rangle$ because of the coherent
population trapping in the $\Lambda$ system. Similarly we can
calculate the population in $|b\rangle$.  Although the unperturbed
EIT condition leads to no steady-state population in $|b\rangle$,
the perturbed EIT will give rise to non-zero population.  As above,
this can be calculated and to third order in $\Omega_3$ we obtain
\begin{eqnarray}
\rho_{bb} = \frac{\Omega_1^2\Omega_2^2}{(\Omega_1^2 + \Omega_2^2)^2}
\frac{\Omega_3^2} {\Delta_3^2 + (\Gamma/2)^2} +
\mathcal{O}[\Omega_3]^4.
\end{eqnarray}
In general, if we have $\Omega_2 = \sqrt{z} \Omega_1$, then we will
have
\begin{eqnarray}
\rho_{bb} = \frac{z}{(z+1)^2}\frac{\Omega_3^2}{\Delta_3^2 +
(\Gamma/2)^2}, \label{eq:Sect3rhobb} \\
\rho_{dd} = \frac{1}{z+1}\frac{\Omega_3^2}{\Delta_3^2 +
(\Gamma/2)^2} \label{eq:Sect3rhocc}.
\end{eqnarray}

Any population occupying the excited states will in general be
available for decoherence and give rise to errors.  By inspection of
Eqs. \ref{eq:Sect3rhobb} and \ref{eq:Sect3rhocc} we observe
qualitatively the same familiar criterion from
Eq.~\ref{eq:rhoddineq} (with minor corrections).

\section{Implications for the design of QND weak nonlinear
detectors}\label{sect:Gates}

Our focus is on the construction of a device capable of achieving a QND measurement of the number of photons in a weak field.  In this section we will combine the previous analyses above with realistic parameters that are achievable using a solid-state slow-light EIT waveguide, which contrasts with the more general discussion of EIT based nonlinear interactions.  Our analysis focusses on QND measurement of a weak field with unknown photon number and we show discrimination between 0, 1 and 2 photons in the unknown field.  In some of the parameter ranges discussed below, we also observe distortions of the probe field.  Whilst this is not a problem for QND discrimination, it may restrict the utility of our discriminator for use in quantum gates \cite{bib:MunroNJP2005}, where the requirement for sequential use of the QND probe favours regimes where the Q-function of the probe is only rotated, and not also distorted by the nonlinear interaction.

We will first describe the appropriate metrics for evaluating the
performance of any such gate in a material independent fashion, and
then conclude by presenting realistic operating conditions in
several potential implementations as a guide for future demonstrations.  For clarity, throughout this section we will restrict ourselves to the case that the $\ket{a}-\ket{b}$ transition is probed by a weak coherent state, and the $\ket{c}-\ket{d}$ transition has the unknown signal field.

One of the most important parameters to determine the strength of
the measurement signal is the effective Rabi frequency of the
unknown pulse. Analysis of this will show that spatially confined
structures (e.g. waveguides) will have a considerable advantage over
free-space implementations. Following \cite{bib:BeausoleilJMO2004} we may express
the single-photon Rabi frequency as
\begin{eqnarray}
\tilde{\Omega}_3 = \frac{\lambda_{12}}{4\pi} \sqrt{3 \Gamma f
\mathcal{N} l} \label{eq:SinglePhotonRabi}
\end{eqnarray}
where we have introduced $\lambda_{12}$, the resonant transition
free-space wavelength, $l$ the length of the medium (or
with imperfect group velocity matching it will be the length of the
effective interaction region) and recall that $f$ is the bandwidth
of the single-photon pulse. The Rabi frequency is then
\begin{eqnarray}
\Omega_3 = \tilde{\Omega}_3 \sqrt{n_3},
\end{eqnarray}
and the other Rabi frequencies can be defined similarly. Note that
there is usually a dependence on the beam waist in
Eq.~\ref{eq:SinglePhotonRabi}, however this is compensated by the
number of interacting four-state systems within the single photon
spot size (so a larger spot interacts with more systems but with
less strength). However, waveguide structures will still have
significant advantages in minimising the beam cross-sectional area
compared with free-space structures, with the effective medium
length in free-space being ultimately limited by the Rayleigh range
of the beam.

We first need to connect the microscopic description presented above
with the macroscopically observable quantities.  In particular, when
considering the four-state system with the probe field on the
$|i\rangle-|j\rangle$ transition, the phase shift seen by the probe
\cite{bib:BeausoleilJMO2004}
\begin{eqnarray}
K_{ij} = \Omega_i\varrho_{ij},
\end{eqnarray}
and the evolution of a state, $\ket{\alpha_i}$, impinging on the
$\ket{i}-\ket{j}$ transition for time $t$ is
\begin{eqnarray}
\ket{\alpha'_i} = \exp\left(i K_{ij} t\right) \ket{\alpha_i}. \label{eq:AlphaEvol}
\end{eqnarray}

Before exploring numerical examples, it is instructive to study the
nonlinear optical processes.  In particular, it is natural to
consider two transitions to be probed to effect quantum
non-demolition measurements, i.e. we could probe the
$\ket{a}-\ket{b}$ transition, with the unknown field on
$\ket{c}-\ket{d}$ transition, or probe the $\ket{c}-\ket{d}$
transition with unknown field on the $\ket{a}-\ket{b}$ transition.
Substituting for the density matrix elements (and assuming
$\Omega_2$ is a classical pump field) we have
\begin{eqnarray}
K_{ab} &=& -\sqrt{\frac{\pi}{2\gamma^2}}\frac{\tilde{\Omega}_1^2
\Omega_2^2 \tilde{\Omega}_3^2 n_1 n_3}{\left(\tilde{\Omega}_1^2 n_1
+ \Omega_2^2\right)^2} J\left(\frac{\Delta_0^2}{2\gamma^2}\right), \\
\mathrm{and,~} K_{cd} &=& -\sqrt{\frac{\pi}{2\gamma^2}}\frac{\tilde{\Omega}_1^2
\tilde{\Omega}_3^2 n_1 n_3}{\left(\tilde{\Omega}_1^2 n_1 +
\Omega_2^2\right)} J\left(\frac{\Delta_0^2}{2\gamma^2}\right).
\end{eqnarray}
which allows the calculation of either phase shift depending on the
relative strengths of the fields.

To understand these nonlinearities it is instructive to consider
certain limits.  One of the common limits is when the pump is strong
and the single photon fields weak, i.e. $\Omega_2 \gg
\tilde{\Omega}_1 \sqrt{n_1}, \tilde{\Omega}_3 \sqrt{n_3}$.  In this
case the population will be optically pumped predominantly into
$\ket{a}$ and the $N$ system reverts to two weakly coupled two-state
transitions. In each case then the phase shift becomes
\begin{eqnarray}
K_{\rm{weak}} = -\sqrt{\frac{\pi}{2\gamma^2}}
\frac{\tilde{\Omega}_1^2\tilde{\Omega}_3^2 n_1
n_3}{\Omega_2^2}J\left(\frac{\Delta_0^2}{2\gamma^2}\right),
\end{eqnarray}
which is an ideal cross-Kerr nonlinearity, but is reduced by the
Rabi frequency of the strong pump.

Another interesting limit is when $\tilde{\Omega}_1^2 n_1 =
\Omega_2$.  Note that this limit will never exactly be reached for a
coherent state probe due to the uncertainty in $n_1$.  In this case
we obtain
\begin{eqnarray}
K_{ab}^{\rm{equal}} =
-\sqrt{\frac{\pi}{2\gamma^2}}\frac{\tilde{\Omega}_3^2n_3}{4}J\left(\frac{\Delta_0^2}{2\gamma^2}\right), \label{eq:KabEqual} \\
K_{cd}^{\rm{equal}} = -\sqrt{\frac{\pi}{2\gamma^2}}
\frac{\tilde{\Omega}_3^2
n_3}{2}J\left(\frac{\Delta_0^2}{2\gamma^2}\right).
\end{eqnarray}
So there is only a factor of two difference between the schemes.  Note that because of the probing condition, we should interpret $K_{ab}$ as the desired cross-Kerr effect to realise QND measurement, and $K_{cd}$ represents a self-Kerr effect, although there are still cross-Kerr nonlinearities at work as the $n_1$ term
has been removed only by the special choice of the limit.

\begin{figure}[tb!]
\includegraphics[width=0.8\columnwidth,clip]{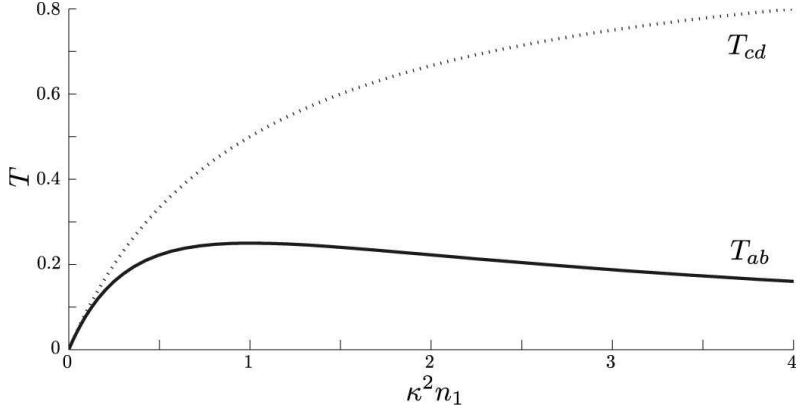}
\caption{\label{fig:KappaScaling} Scalings of the cross-Kerr ($T_{ab}$ - solid lines) and Kerr ($T_{cd}$ - dotted lines)
nonlinearities as a function of $\kappa^2 n_1 = \Omega_1/\Omega_2$  for the case of a weak coherent pulse applied to the $\ket{a}-\ket{b}$ transition and the unknown (possibly single photon) field applied to the $\ket{c}-\ket{d}$ transition.  The maximum of $T_{ab}$ occurs for the case that $\Omega_1 = \Omega_2$.}
\end{figure}

To explore the optimal parameter regime, we define $\kappa \equiv
\tilde{\Omega}_1/\Omega_2$, which gives
\begin{eqnarray}
K_{ab} = -\sqrt{\frac{\pi}{2\gamma^2}} \tilde{\Omega}_3^2 n_3
J\left(\frac{\Delta_0^2}{2\gamma^2}\right)
\frac{\kappa^2 n_1}{\left(\kappa^2 n_1 + 1\right)^2}, \\
K_{cd} = -\sqrt{\frac{\pi}{2\gamma^2}} \tilde{\Omega}_3^2 n_3
J\left(\frac{\Delta_0^2}{2\gamma^2}\right) \frac{\kappa^2
n_1}{\kappa^2 n_1 + 1}.
\end{eqnarray}
To directly compare these results, in Fig.~\ref{fig:KappaScaling} we
show $T_{ab} \equiv \kappa^2 n_1 /(\kappa^2 n_1 + 1)^2$ and
$T_{cd} \equiv \kappa^2 n_1 /(\kappa^2 n_1 + 1)$ as a function of $\kappa^2 n_1$ for varying $n_1$, which is equivalent to the ratio $\Omega_1/\Omega_2$.  The $T$ show us the important regimes and there are many features of interest here.  Firstly we see that $T_{ab}$ is maximised for $\kappa^2 = 1/\sqrt{n_1}$ at $T_{ab} = 1/4$, which is the equal case from from Eq.~\ref{eq:KabEqual}.  This demonstrates  the result anticipated earlier that the largest phase shift is found when the Rabi frequencies of the fields in the EIT system are equal.  Secondly we note that $T_{cd}$ is always greater than $T_{ab}$, and is monotonically increasing.  One must realise, however, that the nature of the nonlinearity is different for the two probing conditions.  By scaling with $\tilde{\Omega_1} n_1$, we are \emph{implicitly} treating the $\ket{a}-\ket{b}$ transition as the probed transition, and $\ket{c}-\ket{d}$ as the unknown transition.  Hence $T_{ab}$ and $T_{cd}$ in this context refer to scalings of the cross-phase and self-phase modulations respectively.  Note that $T_{cd}$ asymptotes to 1 as $\kappa^2 n_1 \rightarrow \infty$.

The correct way to explore the nonlinear effect and hence QND
measurement of the unknown field on the probe is to determine the
Q-function of the state after the interaction. To effect a QND
measurement, we require the Q-functions of the probe beam with and
without a single photon in the channel being monitored to be
distinguishable. Explicitly, returning to Eq.~\ref{eq:AlphaEvol} we take the initial state of the probe
field to be a coherent state $|\alpha\rangle$, with mean photon
number $|\alpha|^2$.  After interacting with this susceptibility,
$K_{ab}$ for a period of time $t$, the probe will be in the state
\begin{eqnarray}
|\alpha'\rangle = \exp\left(-\frac{|\alpha|^2}{2}\right)
 \sum_{n_1=0}^{\infty}\frac{\alpha^{n_1}}{\sqrt{n_1!}} \exp\left(i
 K_{ab} t\right)|n_1\rangle,
\end{eqnarray}
and recall that $K_{ab}$ is a function of both $n_1$ and $n_3$. Note that in general $|\alpha'\rangle$ will \emph{not}
be a coherent state following the interaction. We may define the Q-function for the state
after interacting with the non-linear medium, which is
\begin{eqnarray}
Q(\beta) &=& |\langle \beta|\alpha'\rangle|^2, \\
 &=& \sum_{n_1=0}^{\infty} \frac{\exp\left(-\frac{|\alpha|^2 + |\beta|^2 - 2n_1}{2}\right)}{\sqrt{2\pi
 n_1}}
  \left(\frac{\alpha\beta^*}{n_1}\right)^{n_1}\exp\left(iK_{ab}t\right) \times
  \nonumber \\
 & & \sum_{m_1=0}^{\infty} \frac{\exp\left(-\frac{|\alpha|^2 + |\beta|^2 - 2m_1}{2}\right)}{\sqrt{2\pi
  m_1}}
  \left(\frac{\alpha^*\beta}{m_1}\right)^{m_1}\exp\left(-iK^*_{ab}t\right)
\end{eqnarray}
Note that as required, when $\Omega_3 = 0$ (i.e. the case of
no-photon on the $|c\rangle-|d\rangle$ channel), $K_{ab}=0$ and
the Q-function is simply the same Q-function expected for no
interaction, i.e. there is no self-Kerr modulation.

The candidate system for realising a QND measurement that we are
considering is a monolithic single-mode diamond waveguide (at
$637~\mathrm{nm}$ containing NV$^-$ centres, which is being pursued using a number of different fabrication strategies e.g. \cite{bib:OliveroAdvMat2005,bib:Fairchild,bib:HiscocksDRM2008}.  The energy levels show many possible configurations for achieving the four-state system under consideration \cite{bib:SantoriOptExp2006,bib:TamaratCM} and we will not delve further into these schemes, apart from noting that the transition dipole moments can be achieved and to some extent tuned in situ.  Maximal coupling requires the smallest width single mode waveguides, which for the zero phonon line of NV$^-$ corresponds to a cross section of around $200\times 200~\mathrm{nm}^2$. To avoid potential cross coupling between centres, we require an inter-centre spacing at least around $1~\mu\mathrm{m}$, which for the waveguide under consideration corresponds to an atomic density of $\mathcal{N} = 4\times 10^{20}~\mathrm{m}^{-3}$.  This level is dilute but achievable using ion implantation into ultra-low N synthetic diamond (e.g. Ref.~\cite{bib:MeijerAPL2005}).  Also for NV
diamond, we have $\varepsilon_R \sim 10$ and $\mu_{ab} \sim
10^{-30}$ on the zero-phonon line transition. 

\begin{figure}[tb!]
\includegraphics[width=0.8\columnwidth,clip]{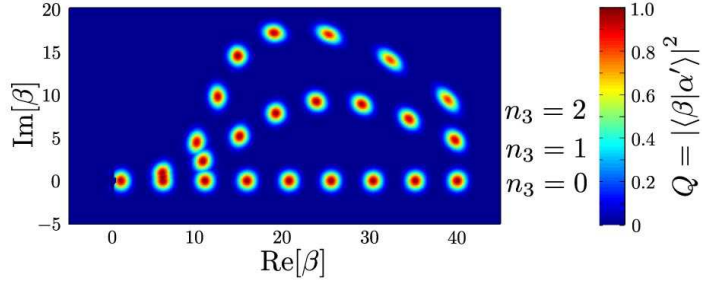}
\caption{\label{fig:QFunc} Q-functions for various conditions
necessary for QND measurements using realistic parameters.  The
three sets of radial distributions correspond to the output probe
beam when the signal has 0, 1 or 2 photons.  Sets of distributions
at constant radius from the origin are grouped with equal
$\ket{\alpha}$.  The increase in effective resolution with
increasing $\alpha$ is demonstrated in this plot, and as discussed
in the text, there is no apparent self-nonlinear processes.
Parameters were chosen to correspond closely to expected EIT
conditions in low N diamond containing NV$^-$ colour centres, and these of reported in Table~\ref{tab:NVParams}.}
\end{figure}

To determine the bandwidth of the pulse to be measured, we rearrange
Eq.~\ref{eq:AbsBandWidth} for an absorption of $\mathcal{A} = 1\%$.
This implies that the bandwidth of probe and single photon field
should be $f=8.9\times 10^{-3} \Gamma = 740~\mathrm{kHz}$, or
approximately $\Gamma/100$.  With this bandwidth, we may immediately
determine the single photon Rabi frequency, which from
Eq.~\ref{eq:SinglePhotonRabi} is $\tilde{\Omega}_3 = 195\mathrm{MHz}$.  The group velocity depends on the EIT condition and also the number of photons in the probe field.

Under these assumptions, and assuming the reduced inhomogeneous
broadening for NVD on the zero phonon line that has been observed in
low N diamond \cite{bib:SantoriOptExp2006,bib:GreentreeJPCM2006,bib:WaldermannDRM2007} of $\gamma =
10~\mathrm{GHz}$, we are now able to fully model the rotation of
the probe field in the QND measurement. Results from simulation are
shown in Fig.~\ref{fig:QFunc} which shows the Q-function for various
values of $\alpha$ and number of photons in the signal beam (field
3).   A full list of parameters used in the calculations are provided in Table~\ref{tab:NVParams}.  

Fig.~\ref{fig:QFunc} shows families of Q-functions for the state of the probe field after traversing the EIT medium.  There are three sets of lines. The lowest, horizontal set corresponds to the case that the signal field (field 3) had no photons, and represents the unperturbed probe field.  This highlights the fact that the three-state EIT system does not exhibit self-phase modulation.  Each point along this set corresponds to increasing $\alpha$ from 1 to 40.  The upper two curves correspond to the probe field after traversing the medium with the signal field in the one photon (middle curve) or two photon (highest curve) state respectively, with $\alpha$ increasing to the right as before.  QND measurement can be inferred whenever the Q-functions of the probe field corresponding to 0, 1 and 2 signal photons do not overlap.  This can be seen clearly in the regime $\alpha >\sim 15$ (fourth set of Q-functions).  Note that the phase shift is not linear in the regimes we are considering, and is greater than would be expected from the simple linear phase shift from an ideal cross-Kerr medium.  This is likely due to the presence of higher-order nonlinearities in the regime $\Omega_1 = \Omega_2$, although we have not made a detailed study of them.  The distinguishability is maximised at around $\alpha \sim 25$ for these parameters.  Note that this is earlier than the maximum expected from $T_{ab}$, which we attribute to group velocity dispersion reducing the effective interaction time for certain modes of the probe field.  A complete study of these processes is beyond the scope of this work.  A further point to note is that although QND discrimination can be made with very high certainty for $\alpha >\sim 14$, there is some noticeable distortion, especially in the 2-signal photon branch.  This indicates that further work is required before strong claims of the effectiveness of this scheme for quantum gate operation can be made.

\begin{table}[htdp]
\caption{Parameters for NV diamond waveguide system under consideration as QND gate}
\begin{center}
\begin{tabular}{|c|c|}
\hline Waveguide dimensions & $w_0 \times w_0 \times l = 200~\mathrm{nm} \times 200~\mathrm{nm} \times 200~\mu\mathrm{m}$  \\
\hline Atomic concentration & $\mathcal{N} = 4\times 10^{20}~\mathrm{m}^{-3}$ \\
\hline Dipole moment (ZPL) & $\mu_{ab} = 10^{-30}~\mathrm{Cm}^{-1}$  \\
\hline Transition frequency & $\lambda = 637~\mathrm{nm}$ \\
\hline Relative permittivity of diamond & $\epsilon_R = 10$ \\
\hline Homogeneous linewidth & $\Gamma = 83~\mathrm{MHz}$ \\
\hline Inhomogeneous linewidth & $\gamma = 10~\mathrm{GHz}$ \\
\hline Pulse bandwidth & $f = 740~\mathrm{kHz}$ \\
\hline Semiclassical Rabi frequency & $\Omega_2 = \Gamma/10$ \\
\hline Field ratios & $\kappa = \tilde{\Omega}_1/\Omega_2 = 1/50$ \\
\hline Signal per photon Rabi frequency & $\tilde{\Omega}_3 = \frac{\lambda}{4\pi}\sqrt{3\Gamma f \mathcal{N} l} = 195~\mathrm{MHz}$ \\
\hline Detuning scaling & $J = 1/6$ \\
\hline Group velocity & $v_{g_{ab}} = \frac{c \varepsilon_0 \varepsilon_R \hbar \Omega_2^2}{\omega \mathcal{N} \mu_{ab}^2}\left(\kappa^2 n_1 + 1\right) = 2.88 \times 10^4 \left(\frac{n_1}{2500}+1\right)~\mathrm{ms}^{-1}$ \\ \hline
\end{tabular}
\end{center}
\label{tab:NVParams}
\end{table}%

The parameters used in the equations used to generate the results
shown in Fig.~\ref{fig:QFunc} are fairly typical of what has already
been achieved or will be soon achieved in NVD.  With variations in
the parameters, there will be changes in the resulting phase shifts,
but the parameters that we chose were not ``fine tuned". 

\section{Conclusions}
We have performed investigations of three-state EIT in an
inhomogeneously broadened sample with the aim of determining the
necessary conditions for performing a quantum non-demolition
measurement in a realistic solid-state medium, diamond containing
the nitrogen-vacancy colour centre. Our results suggest that even in
the presence of the relatively large inhomogeneous linewidth of
these systems, QND measurements are possible using relatively modest
extensions of the existing state of the art (demonstration of single
mode waveguides of order $200~\mu\mathrm{m}$ long, and group
velocity compensated propagation).  These conclusions add
substantial impetus to the ongoing push for diamond based quantum
photonics and offer increased support for quantum computing based on
weak nonlinear interactions.

\ack

ADG would like to thank the Japan Society for the Promotion of Science for an
Invitation Fellowship that facilitated some of this work and is the recipient of an
Australian Research Council Queen Elizabeth II Fellowship (DP0880466). LCLH is
the recipient of an Australian Research Council Australian Professorial Fellowship
(DP0770715) This work was supported in part by MEXT, NICT and JSPS in Japan
and the EU project QAP. This work was partially supported by DARPA and the Air
Force Office of Scientific Research through AFOSR Contract No. FA9550-07-C-0030,
and in part was supported by the Australian Research Council Centre of Excellence
for Quantum Computer Technology, and by the DEST.



\section*{References}
\begin{thebibliography}{99}
\bibitem{bib:DowlingQP02} Dowling JP and Milburn GJ 2003 {\it Philosophical Transactions of the Royal Society of London Series A: Mathematical, Physical and Engineering Sciences} {\bf 361}, 1655

\bibitem{bib:HarrisPRL1990} Harris SE, Field JE and Imamo\u{g}lu A
1990 \PRL {\bf 64}, 1107

\bibitem{bib:BollerPRL1991}Boller K-J, Imamo\u{g}lu A, and Harris SE 1991 \PRL {\bf 66}, 2593.

\bibitem{bib:ScullyPRL1992} Scully MO and Fleischhauer M 1992 \PRL
{\bf 69}, 1360

\bibitem{bib:MerriamOL1999} Merriam AJ, Sharpe SJ, Xia H, Manuszak
D, Yin GY and Harris SE 1999 {\it Opt. Lett.} {\bf 24}, 625

\bibitem{bib:DormanPRA2000} Dorman C, Kucukkara I and Marangos JP 2000
{\it Phys. Rev. A}, {\bf 61} 013802

\bibitem{bib:HarrisPRL1998} Harris SE and Yamamoto 1998 \PRL {\bf
81}, 3611

\bibitem{bib:Beausoleil2004} R. G. Beausoleil, W. J. Munro, D. A. Rodrigues, and T. P.
Spiller 2004 {\it J. Mod. Opt.} {\bf 51}, 2441.

\bibitem{bib:TuckerJLT2005} Tucker RS, Ku PC and Chang-Hasnain CJ 2005 {\it
Journal of Lightwave Technology} {\bf 23}, 4046

\bibitem{bib:MatskoPRA2007} Matsko AB, Strekalov D, Savchenkov AA and
Maleki L 2007 {\it Phys. Rev. A} {\bf 76} 013806

\bibitem{bib:ZibrovPRL1995} Zibrov AS, Lukin MD, Hollberg L, Nikonov
DE, Scully MO, Robinson HG and Velichansky VL 1996 \PRL {\bf 76},
3935

\bibitem{bib:HamOC1997} Ham BS, Hemmer PR and Shahriar MS 1997 {\it
Optics Communications} {\bf 144}, 227

\bibitem{bib:WeiJOB1999} Wei C and Manson NB 1999 \JOB {\bf 1}, 464

\bibitem{bib:HemmerOptLett2001} Hemmer PR, Turukhin A, Shahriar MS and Musser J 2001 {\it Opt. Lett.} {\bf 26}, 361

\bibitem{bib:KuznetsovaPRA2002} Kuznetsova E, Kocharovskaya O, Hemmer P and Scully MO 2002 {\it Phys. Rev. A} {\bf 66}, 063802

\bibitem{bib:SantoriOptExp2006} Santori C, Fattal D, Spillane SM,
Fiorentino M, Beausoleil RG, Greentree AD, Olivero P, Draganski M,
Rabeau JR, Reichart P, Gibson BC, Rubanov S, Huntington ST,
Jamieson~DN and Prawer S 2006 {\it Optics Express} {\bf 14}, 7986

\bibitem{bib:SantoriPRL2006} Santori C, Tamarat P, Neumann P,
Wrachtrup J, Fattal D, Beausoleil R, Rabeau J, Olivero P, Greentree
AD, Prawer S, Jelezko F and Hemmer P 2006 \PRL {\bf 97}, 247401

\bibitem{bib:AlegrePRB2007} Mayer Alegre TP, Santori C, Medeiros-Ribeiro G and
Beausoleil RG 2007 {\it Phys. Rev. B} {\bf 76}, 165205

\bibitem{bib:Chen1998} Chen HX, Durrant AV, Marangos JP and Vaccaro
JA 1998 {\it Phys. Rev. A} {\bf 58}, 1545

\bibitem{bib:HauNature1999} Hau LV, Harris SE, Dutton Z and Behroozi
CH 1999, Nature (London) {\bf 397}, 594

\bibitem{bib:ImotoPRA1985} Imoto N, Haus HA and Yamamoto Y 1985 {\it
Phys. Rev. A} {\bf 32}, 2287

\bibitem{bib:MunroPRA2005} Munro WJ, Nemoto K, Beausoleil RG and Spiller TP
2005 {\it Phys. Rev. A} {\bf 71}, 033819

\bibitem{bib:NemotoPLA2005} Nemoto K and Munro WJ 2005 \textit{Phys. Lett.
A} {\bf 344}, 104

\bibitem{bib:NemotoPRL2004} Nemoto K and Munro WJ 2004, \PRL {\bf
93}, 250502

\bibitem{bib:MunroNJP2005} Munro WJ, Nemoto K and Spiller TP 2005
\NJP {\bf 7}, 137

\bibitem{bib:MunroJOptB2005} Munro WJ, Nemoto K, Spiller TP, Barrett
SD, Kok P and Beausoleil RG 2005 \JOB {\bf S135}

\bibitem{bib:SpillerNJP2006} Spiller TP, Nemoto K, Braustein SL,
Munro WJ, van Loock P and Milburn GJ 2006, \NJP {\bf 8}, 30

\bibitem{bib:LoockPRL2006} van Loock P, Ladd TD, Sanaka K, Yamaguchi
F, Nemoto K, Munro WJ and Yamamoto Y 2006, \PRL {\bf 96}, 240501

\bibitem{bib:LouisNJP2007} Louis SGR, Nemoto K, Munro WJ and Spiller
TP 2007, \NJP {\bf 9} 193

\bibitem{bib:KraussJPD2007} Krauss TF 2007, \JPD {\bf 40}, 2666

\bibitem{bib:AltugAPL2005} Altug H and Vu\v{c}kovic J 2005, {\it Appl. Phys. Lett.} {\bf 86}, 111102 

\bibitem{bib:FleischhauerRMP2005} Fleischhauer M, Imamoglu A and
Marangos JP 2005 \RMP {\bf 77}, 633

\bibitem{bib:SchmidtOL1996} Schmidt H and Imamoglu A 1996 \textit{Opt. Lett.}
{\bf 21}, 1936

\bibitem{bib:KashPRL1999} Kash MM, Sautenkov VA, Zibrov AS, Hollberg
L, Welch GR, Lukin MD, Rostovtsev Y, Fry ES and Scully MO 1999 \PRL
{\bf 82} 5229

\bibitem{bib:TurukhinPRL2002} Turukhin AV, Sudarshanam VS,
Shahriar MS, Musser JA, Ham BS and Hemmer PR 2002 \PRL {\bf 88}
023602


\bibitem{bib:ImamogluPRL1997} Imamo\u{g}lu A, Schmidt H, Woods G and
Deutsch M 1997 \PRL {\bf 79}, 1467

\bibitem{bib:RebicJOB1999} Rebi\'{c} S, Tan SM, Parkins AS and Walls DF
1999 \JOB {\bf 1}, 1490

\bibitem{bib:WernerPRA2000} Werner MJ and Imamo\u{g}lu A 2000 {\it Phys.
Rev. A} {\bf 61}, R011801

\bibitem{bib:GreentreeJOB2000} Greentree AD, Vaccaro JA, de Echaniz SR,
Durrant AV and Marangos JP 2000 \JOB {\bf 2} 252

\bibitem{bib:HartmannQP2006} Hartmann MJ, Brand\~{a}o FGSL
and Plenio MB, 2006 {\it Nature Physics} {\bf 2} 849

\bibitem{bib:BermelPRA2006} Bermel P, Rodriguez A, Johnson SG,
Joannopoulos JD and Solja\v{c}i\'{c} M 2006 {\it Phys. Rev. A} {\bf
74} 043818

\bibitem{bib:BirnbaumNat2005} Birnbaum KM, Boca A, Miller R, Boozer AD,
Northup TE and Kimble HJ 2005 {\it Nature (London)} {\bf 436} 87

\bibitem{bib:YanPRA2001} Yan M, Rickey EG and Zhu Y 2001 {\it Phys.
Rev. A} {\bf 64} 013412

\bibitem{bib:EchanizPRA2001} de Echaniz SR, Greentree AD, Durrant
AV, Segal DM, Marangos JP and Vaccaro JA 2001 {\it Phys. Rev. A}
{\bf 64} 013812

\bibitem{bib:PaspalakisJMO2002} Paspalakis E and Knight PL 2002 {\it
J. Mod. Opt} {\bf 49} 87

\bibitem{bib:Zubairy2002} Zubairy MS, Matsko AB and Scully MO 2002
{\it Phys. Rev. A} {\bf 65} 043804

\bibitem{bib:GreentreePRA2003} Greentree AD, Richards D, Vaccaro JA,
Durrant AV, de Echaniz SR, Segal DM and Marangos JP 2003 {\it Phys.
Rev. A} {\bf 67} 023818

\bibitem{bib:RebicPRA2004} Rebi\'{c} S, Vitali D, Ottaviani C, Tombesi P,
Artoni M, Cataliotti F and Corbal\'{a}n R 2004 {\it Phys. Rev. A}
{\bf 70} 032317

\bibitem{bib:LukinPRL2000} Lukin MD and Imamo\u{g}lu A 2000 \PRL
{\bf 84} 1419

\bibitem{bib:WangPRL2006} Wang ZB, Marzlin KP and Sanders BC 2006
\PRL {\bf 97} 063901

\bibitem{bib:VlasovPRE1999} Vlasov YA, Petit S, Klein G, H\"{o}nerlage B and
Hirlimann Ch 1999 {\it Phys. Rev. E} {\bf 60} 1030

\bibitem{bib:VlasovNature2005} Vlasov YA, O'Boyle M, Hamann HF and
McNab SJ 2005 {\it Nature} {\bf 483} 65

\bibitem{bib:BeveratosEPJD2002} Beveratos A, K\"{u}hn S, Brouri R, Gacoin
T, Poizat J-P and Grangier P 2002 {\it Eur. Phys. J. D} {\bf 18} 191

\bibitem{bib:JelezkoPRL2004} Jelezko F, Gaebel T, Popa I, Gruber A
and Wrachtrup J 2004 \PRL {\bf 92} 076401

\bibitem{bib:JelezkoPRL2004b} Jelezko F, Gaebel T, Popa I,
Domhan M, Gruber A and Wrachtrup J 2004 \PRL {\bf 93} 130501

\bibitem{bib:GaebelNatPhys2006} Gaebel T, Domhan M, Popa I, Wittmann
C, Neumann P, Jelezko F, Rabeau JR, Stavrias N, Greentree AD, Prawer
S, Meijer J, Twamley J, Hemmer PR and Wrachtrup J, {\it Nature
Physics} {\bf 2} 408

\bibitem{bib:HansonPRL2006} Hanson R, Mendoza FM, Epstein RJ and
Awschalom DD 2006 \PRL {\bf 97} 087601

\bibitem{bib:DuttScience2007}  Dutt MVG, Childress L, Jiang L, Togan
E, Maze J, Jelezko F, Zibrov AS, Hemmer PR and Lukin MD 2007 {\it
Science} {\bf 316} 1312

\bibitem{bib:WangAPL2007} Wang CF, Choi YS, Lee JC, Hu EL, Yang J and
Butler JE 2007 {\it Appl. Phys. Lett.} {\bf 90}, 081110

\bibitem{bib:OliveroAdvMat2005} Olivero P, Rubanov S, Reichart P,
Gibson B, Huntington S, Rabeau J, Greentree AD, Salzman J, Moore D,
Jamieson DN and Prawer S 2005 {\it Advanced Materials} {\bf 17},
2427

\bibitem{bib:GreentreeJPCM2006} Greentree AD, Olivero P, Draganski
M, Trajkov E, Rabeau JR, Reichart P, Gibson BC, Rubanov S,
Huntington ST, Jamieson DN and Prawer S, \JPCM {\bf 18}, S825

\bibitem{bib:Fairchild} Fairchild BA, Olivero P, Rubanov S, Greentree AD,
Waldermann F, Taylor RA, Walmsley I, Smith JM, Huntington S,
Gibson BC, Jamieson DN, and Prawer S 2008 {\it Advanced Materials} {\bf 20}, 4793.

\bibitem{bib:RabeauAPL2005} Rabeau JR, Huntington ST, Greentree AD
and Prawer S, {\it Appl. Phys. Lett.} {\bf 86}, 134104

\bibitem{bib:Redman} Redman D, Brown S and Rand SC 1992
{\it J. Opt. Soc. Am. B} \textbf{9}, 768

\bibitem{bib:TamaratPRL2006} Tamarat Ph, Gaebel T, Rabeau JR, Khan M,
Greentree AD, Wilson H, Hollenberg LCL, Prawer S, Hemmer P, Jelezko
F and Wrachtrup J 2006 \PRL {\bf 97}, 083002

\bibitem{bib:TamaratCM} Tamarat Ph, Manson NB, Harrison JP, McMurtrie RL, Nizovtsev A, Santori C, Beausoleil RG, Neumann P, Gaebel T, Jelezko F, Hemmer P and Wrachtrup J 2008 \NJP {\bf 10}, 045004


\bibitem{bib:ShoreBook} B. W. Shore, {\it The Theory of Coherent Atomic
Excitation, Volume 2, Multilevel Atoms and Incoherence} (John Wiley
and Sons, New York, 1990).

\bibitem{bib:WielandyPRA1998} Wielandy S and Gaeta AL 1998
{\it Phys. Rev. A} {\bf 58}, 2500.

\bibitem{bib:MullerPRA2000} M\"{u}ller M, Homann F, 
Rinkleff R-H, Wicht A, and Danzmann K 2000 {\it Phys. Rev. A}  {\bf 62},
060501(R).

\bibitem{bib:GeaBanalochePRA1995} Gea-Banaloche J, Li Y, Jin S,
and Xiao M 1995 {\it Phys. Rev. A} {\bf 51}, 576.

\bibitem{bib:VemuriPRA1996} Vemuri G and Agarwal GS 1996 {\it Phys. Rev.
A} {\bf 53}, 1060.

\bibitem{bib:AllenBook} Allen L and Eberly J H 1975 Optical Resonance and Two Level Systems (Chichester: Wiley)

\bibitem{bib:Rohde2008} Rohde PP, Munro WJ, Ralph TC, van Loock P
and Nemoto K 2008 {\it Quantum Information and Computation} {\bf 8},
0053.

\bibitem{bib:Kok2007} Kok P 2008 {\it Phys. Rev. A} {\bf 77}, 013808.

\bibitem{bib:BeausoleilJMO2004} Beausoleil RG, Munro WJ, Spiller TP
2004 {\it J. Mod. Opt.} {\bf 51} 1559.

\bibitem{bib:HiscocksDRM2008} Hiscocks MP, Kaalund CJ, Ladouceur F, Huntington ST, Gibson BC, 
Trpkovski S, Simpson D, Ampem-Lassen E, Prawer S, Butler JE 2008 {\it Diamond and Related Materials} {\bf 17}, 1831. 

\bibitem{bib:MeijerAPL2005}  Meijer J, Burchard B, Domhan M, Wittmann C, Gaebel T, Popa I, Jelezko F and Wrachtrup J 2005 {\it Appl. Phys. Lett.} {\bf 87}, 261909 

\bibitem{bib:WaldermannDRM2007} Waldermann FC, Olivero P, Nunn J, Surmacz K, Wang ZY, Jaksch D, Taylor RA, 
Walmsley IA, Draganski M, Reichart P, Greentree AD, Jamieson DN, Prawer S 2007 {\it Diamond and Related Materials} {\bf 16}, 1887 

\end{thebibliography}
\end{document}